\documentclass[usenatbib]{mn2e}

\pdfoutput=1

\usepackage[pdftex]{graphicx} 

\usepackage{times}

\usepackage{rotate} 

\usepackage{amssymb} 

\usepackage{rotating}

\usepackage{epstopdf} 

\usepackage{graphicx}

\usepackage{array}

\title[Negative luminosity dependence of the AGN clustering]{The clustering amplitude of X-ray selected AGN at z$\sim$0.8: evidence for a negative dependence on accretion luminosity}

\author[Mountrichas et al.]{G. Mountrichas$^1$,
  A. Georgakakis$^{2,1}$, M.-L. Menzel$^2$, N. Fanidakis$^3$, A. Merloni$^2$,
  \newauthor  Z. Liu$^{2,4}$, M. Salvato$^2$, K. Nandra$^2$\\ \\ 
$^1$National Observatory of Athens, V.  Paulou  \& I.  Metaxa, 11532,  Greece\\ 
$^2$Max Planck Institut f\"{u}r Extraterrestrische  Physik, Giessenbachstra\ss e, 85748 Garching, Germany\\ 
$^3$Max Planck Institut f\"ur Astronomie, K\"onigstuhl, 17
D-69117, Heidelberg, Germany\\
$^4$National Astronomical Observatories, Chinese Academy of Sciences, Beijing 100012, People's Republic of China.}

\begin{document}
\maketitle
\label{firstpage}

\begin{abstract}  

The northern  tile of the  wide-area and shallow XMM-XXL  X-ray survey
field is  used to estimate the  average dark matter halo  
  mass of relatively luminous X-ray  selected AGN [$\rm
  log\,  L_X  (\rm   2-10\,keV)=  43.6^{+0.4}_{-0.4}\,erg/s$]  in  the
redshift interval  $z=0.5-1.2$.  Spectroscopic  follow-up observations
of  X-ray sources  in the  XMM-XXL field  by the  Sloan telescope  are
combined with the VIPERS spectroscopic  galaxy survey to determine the
cross-correlation  signal between  X-ray  selected AGN
(total  of 318)  and galaxies  (about 20,\,000).   We model  the large
scales (2-25\,Mpc)  of the correlation  function to infer a  mean dark
matter halo mass  of $\log M / (M_{\odot} \,  h^{-1}) = 12.50 ^{+0.22}
_{-0.30}$ for the X-ray selected  AGN sample.  This
measurement  is about  0.5\,dex  lower compared  to  estimates in  the
literature of  the mean dark matter  halo masses of
moderate luminosity  X-ray AGN [$L_X (\rm  2-10\,keV)\approx 10^{42} -
  10^{43}\,erg/s$] at  similar redshifts. Our analysis  also links the
mean clustering  properties of moderate  luminosity AGN with  those of
powerful  UV/optically selected  QSOs,  which are  typically found  in
halos with masses few  times $10^{12}\,M_{\odot}$.  There is therefore
evidence   for   a  negative   luminosity   dependence   of  the   AGN
clustering. This is consistent with  suggestions that AGN have a broad
dark matter halo mass distribution with  a high mass tail that becomes
sub-dominant at  high accretion luminosities.  We  further show that
our results are in qualitative  agreement with semi-analytic models of
galaxy  and AGN  evolution, which  attribute  the wide  range of  dark
matter halo  masses among the  AGN population to  different triggering
mechanisms and/or black hole fueling modes.

\end{abstract}


\section{Introduction}

The  study of  the large  scale clustering  of Active  Galactic Nuclei
(AGN)  links accretion  events onto  Supermassive Black  Holes (SMBHs)
with  their  environments  and  provides  useful  constraints  on  the
conditions under  which such events  occur in the  Universe. Different
clustering  properties  are predicted  for  the  AGN depending,  among
others, on  the mechanism  that triggers the  accretion onto  the SMBH
\citep[secular  evolution   vs  mergers,][]{Hopkins2008a,  Bonoli2009,
Bournaud2011},  the origin  of  the accreted  material (e.g.   stellar
winds,  \citealt{Ciotti2007};  hot  atmosphere,  \citealt{Croton2006};
galactic  disk,  \citealt{Fanidakis2011}),   assumptions  on  the  AGN
light-curve  \citep[e.g.][]{Kauffmann2002,  Wyithe2003,  Hopkins2007},
the significance of AGN feedback processes \citep[e.g.][]{Thacker2009,
Fanidakis2013a}.

The different model assumptions above  can be tested by estimating the
clustering of  AGN populations as  a function of redshift  and/or physical
AGN  parameters,  such as  accretion  luminosity,  Eddington ratio  or
black-hole mass.   UV/optically selected broad-line  QSOs for example,
are shown  to live in Dark  Matter Haloes (DMH) with  masses few times
$10^{12}\,{\rm h}^{-1}\,M_{\odot}$ independent  of redshift and with a
weak,       if        any,       accretion-luminosity       dependence
\citep[e.g.][]{Croom2005, Myers2007, daAngela2008, Ross2009, Shen2009,
Shen2013}.   These results  have been  interpreted in  the  context of
major-merger scenarios  for the triggering  of the accretion  onto the
SMBH \citep[e.g.][]{Hopkins2007}.   The weak luminosity  dependence of
the  QSO  clustering also  argues  against  a  tight relation  between
instantaneous accretion luminosity and  black-hole or DMH mass and has
implications   for   the   form   of   the   accretion   light   curve
\citep[e.g.][]{Hopkins2007, Lidz2006}.

Despite  the  significance of  these  results  it  is recognised  that
powerful UV/optically  bright QSOs represent  a small fraction  of the
AGN  population  and  are  biased  against  even  moderately  obscured
sources.  It is therefore  necessary to  complement the  results above
with clustering investigations of  AGN samples selected by alternative
means.   X-ray wavelengths  are advantageous  in that  respect \citep[e.g][]{Brandt_Alexander2015}.  X-ray 
photons,  especially  at  energies  above  the keV  level,  are  least
affected by  obscuring dust  and gas clouds  along the line  of sight.
Also,  the  X-ray  emission   associated  with  stellar  processes  is
typically orders  of magnitude fainter  than the AGN  radiative output
and therefore  dilution effects by  the host galaxy are  negligible at
X-rays even  for low  accretion luminosities. X-ray  surveys therefore
provide  least biased  AGN samples  over a  wide  accretion luminosity
baseline.

Clustering studies of moderate luminosity X-ray selected AGN typically
measure  mean dark  matter halo  masses  of $\approx  10^{13} \,  {\rm
h}^{-1}     \,    M_{\odot}$,     at     least    to     $z\approx1.5$
\citep[e.g.][]{Coil2009,  Krumpe2010,  Allevato2011,  Mountrichas2012,
Mountrichas2013}.   When compared  to  UV/optically selected  powerful
QSOs, this result suggests a negative luminosity dependence of the AGN
clustering,  i.e.    decreasing  mean  dark  matter   halo  mass  with
increasing  accretion  luminosity.   Possible interpretations  include
different  AGN triggering mechanisms  \citep[e.g.][]{Allevato2011}, or
diverse accretion modes  \cite[e.g.][]{Fanidakis2013a} that dominate at
different AGN  luminosity regimes. There are also  attempts to explore
the luminosity  dependence of the clustering using  X-ray selected AGN
only.  These however,  are met with moderate  success.  There are
claims  for  a  positive  correlation  between  X-ray  luminosity  and
clustering but the statistical  significance of those results is small
\citep[e.g.][]{Krumpe2012, Koutoulidis2013}.   A serious limitation is
that  current   X-ray  AGN  samples  are  typically   small  in  size.
Statistical  uncertainties   therefore  dominate  and   do  not  allow
clustering  investigations   over  a  sufficiently   large  luminosity
baseline.   This  issue  can  be  addressed  by  combining  clustering
constraints from X-ray samples  with different depths and survey areas
that  probe different  parts of  the  luminosity function  at a  given
redshift.   There  are currently  numerous  X-ray  surveys with  sizes
typically  $\rm   \la  2\,deg^2$  that  provide   constraints  on  the
statistical  properties  of  moderate  and  low  luminosity  AGN.   In
contrast,  the   number  of  wide-area  X-ray   samples  that  provide
sufficient  statistics  at  the  bright-end of  the  X-ray  luminosity
function is still limited.

One  of the  widest contiguous  X-ray surveys  that has  recently been
completed  is the  XMM-XXL (PI:  Pierre). With  a total  area  of $\rm
50\,deg^2$ on  the sky split  into two equally-sized  sub-regions, the
XMM-XXL is 1-2\,dex larger in size compared to any current pencil-beam
and  deep survey  used  for clustering  investigations.  It  therefore
provides a  unique resource for studying the  population properties of X-ray AGN  that are  underrepresented in  current small-area
X-ray  samples. In this paper we  focus the  clustering  analysis  to the  equatorial
sub-region  of  the  XMM-XXL  field,  which  benefits  from  extensive
follow-up  spectroscopy  of X-ray  sources  as  part  of the  SDSS-III
\citep{Eisenstein2011}  BOSS  \citep[Baryon Oscillation  Spectroscopic
Survey;][]{Dawson2013}    ancillary   observations    programme.    An
additional  advantage  of the  equatorial  XMM-XXL  field  is that  it
overlaps   with   the   VIPERS   \citep[Vimos   Public   Extragalactic
Survey;][]{Guzzo2014}  galaxy  spectroscopic  survey.  This  programme
provides redshifts for about 20,000 galaxies in the interval $0.5\la
z \la 1.2$ within the XMM-XXL field. The Sloan and VIPERS spectroscopy
are  combined to determine  the AGN/galaxy  cross-correlation function
and then  infer the bias and  dark matter halo mass  of relatively luminous X-ray
AGN  [$\rm
  log\,  L_X  (\rm   2-10\,keV)=  43.6^{+0.4}_{-0.4}\,erg/s$]. This approach  is advantageous  for clustering  measurements of
sparse  samples,  such as  AGN,  because  the  statistical errors  are
smaller   compared  to  the   auto-correlation  function.    We  adopt
$\Omega_m=0.3$  $\Omega_{\Lambda}=0.7$  and  $\sigma_8=0.8$.  For  the
clustering    analysis    the    hubble    constant    is    set    to
$H_{0}=100$\,km\,s$^{-1}$Mpc$^{-1}$ and all relevant quantities, such halo
masses, are  parametrised by $h=H_0/100$.   In the calculation  of the
X-ray     luminosities    we     fix  $H_0=70$\,km\,s$^{-1}$\,Mpc$^{-1}$
(i.e. $h=0.7$). This is to allow comparison with previous studies that
also follow similar conventions.

\begin{figure*}
\begin{center}
\includegraphics[height=0.9\columnwidth]{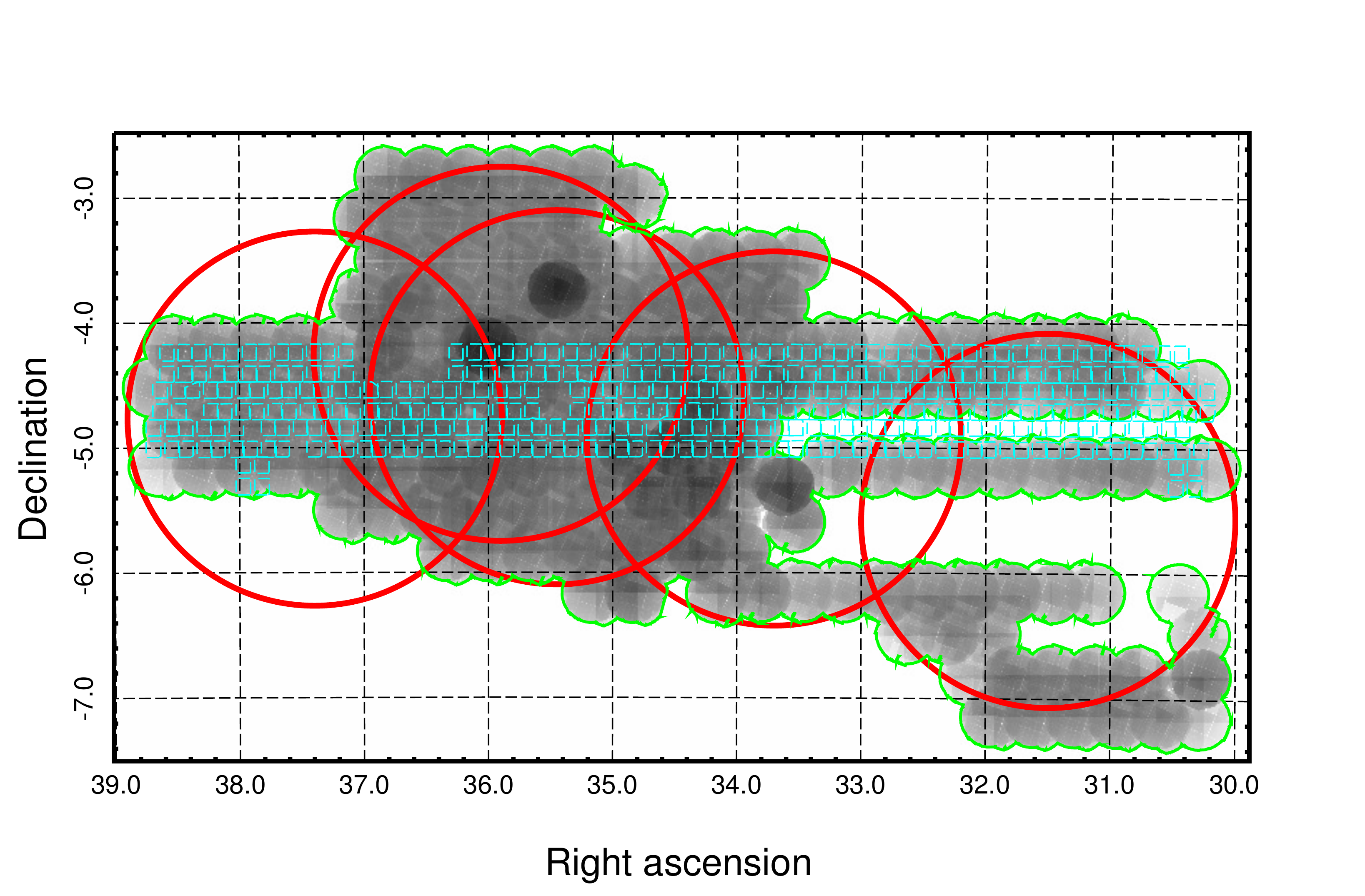}
\end{center}
\caption{Layout  of the  XMM  observations in  the equatorial  XMM-XXL
field used in  this paper. The distribution on  sky coordinates of the
EPIC PN+MOS  exposure maps of  individual XMM pointings is  shown. The
shading  corresponds to  different  exposure times, with white  being
zero. The green line demarcates the  limits of the area covered by the
XMM  data  observed  prior  to  23 January  2012.   XMM-XXL  pointings
observed after this date (total  area of about $\rm 5\,deg^2$) are not
included  in the  analysis.   This incompleteness  in  the X-ray  data
coverage is  manifested by the white (unexposed)  horizontal stripes at
e.g.  $\delta \rm \approx -5.5$\,deg and $\alpha < 33.5$\,deg. The red
circles mark the positions  of the five SDSS-III special spectroscopic
plates used  to target X-ray  sources in the  field. The size  of each
circle  is  3\,deg  in  diameter,  the  field--of--view  of  the  SDSS
spectroscopic plates.  The small cyan squares mark the positions of the
VIMOS field-of-view quadrants of the VIPERS DR1.}
\label{fig_layout}
\end{figure*}

\section{The data} \label{section:samples}







\subsection{Optically selected galaxy sample} \label{section:galaxies}

The   galaxy  spectroscopic   sample   is  from   the  VIPERS   survey
\citep{Guzzo2014, Garilli2014}. In brief the VIPERS programme uses the
VIMOS   \citep[VIsible  MultiObject   Spectrograph,][]{LeFevre2003}  to
perfom deep  optical spectroscopy over 24\,deg$^2$  split between the
W1  and W4  wide tiles  of the  Canada-France-Hawaii  Telescope Legacy
Survey. Potential targets for follow-up spectroscopy are selected to
the  magnitude limit  $i'_{AB}=22.5$  from the  T0006  data
release  of the  CFHTLS  photometric catalogues\footnote{Based on observations obtained with MegaPrime/MegaCam, a joint project of CFHT and CEA/DAPNIA, at the Canada-France-Hawaii Telescope (CFHT) which is operated by the National Research Council (NRC) of Canada, the Institut National des Sciences de l'Univers of the Centre National de la Recherche Scientifique (CNRS) of France, and the University of Hawaii. This work is based in part on data products produced at TERAPIX and the Canadian Astronomy Data Centre as part of the Canada-France-Hawaii Telescope Legacy Survey, a collaborative project of NRC and CNRS.}. The science  motivation of  the VIPERS  is the  study  of the
large-scale  structure   of  the  Universe  and   the  measurement  of
cosmological   parameters  at   redshifts  $z\approx1$.    An  optical
photometric     colour    pre-selection    is     therefore    applied
[$(r-i)>0.5(u-g)$  or $(r-i)>0.7$]  to exclude  low  redshift galaxies
from  the  target sample  and  maximise  the  number of  spectroscopic
identifications  in  the  redshift  interval $z=0.5-1.2$.   The  first
public data release  (PDR-1) of the VIPERS programme  includes a total
of 62,\,862 spectra observed by the  VIMOS instrument in both the W1 and W2
fields \citep{Garilli2014}.  Each spectrum  is assigned a quality flag
that  quantifies  the  reliability  of  the  measured  redshift.   The
analysis presented in this paper uses only the CFHTLS-W1 VIPERS field,
which  has X-ray  coverage from  the XMM-XXL  survey (see  below).  We
select  only galaxies  with flags  2 to  9 inclusive.   This  yields a
sample of 20,109 galaxies in the redshift range $0.5<z<1.2$.

The VIPERS observations strategy is to cover the survey area with only
one  pass  to maximise  the  volume  probed.   This approach  however,
imprints  a characteristic cross  shape on  the survey  footprint (see Fig.
\ref{fig_layout})  as a  result  of the  VIMOS  field-of-view that  is
composed of  four CCDs  with gaps among  them. This pattern  should be
accounted for in large scale clustering investigations.  Additionally,
the maximum number  of science slits on VIMOS  and the surface density
of the targeted  population result in a slit  assignment efficiency of
about  $45\%$  on  the   average,  but  variable  across  the  survey
area. Weather  and night-sky  conditions during the  observations also
affect the  quality of the spectra  and therefore the  success rate of
measuring  reliable  redshifts across  the  survey  area.  The  above
selection and instrumental effects  are accounted for, following the methods
described in \cite{delaTorre2013}. The  fraction of targets which have
a measured spectrum  is defined as the Target  Sampling Rate (TSR) and
the fraction of observed spectra with reliable redshift measurement as
the  Spectroscopic  Sampling Rate  (SSR).   The  latter is  determined
empirically as the ratio between  the number of reliable redshifts and
the total  number of observed  spectra. These two factors  correct for
incompleteness at large  scales and are taken into  account in form of
weights  applied  to  individual  sources when  measuring  the  galaxy
auto-correlation   function   or   the  AGN/galaxy   cross-correlation
function. Incompleteness  at small scales  is also important.  This is
because of missing  small-scale  angular  pairs  due to  the  slit
collisions and the finite number of VIMOS slits. This can be corrected
for by  comparing the angular auto-correlation function  of the target
and spectroscopic samples \citep{delaTorre2013}.

For the correlation measurements,  we also generate a random catalogue
that consists of $\approx 20\times$ the number of galaxies. The random
catalogue  is passed through  both the  photometric mask  (i.e. bright
stars) and the spectroscopic mask (Field-of-View of the spectrograph),
so the random points have the  same sky footprint as the real data. To
test if the  random sample is large enough to  minimise the shot noise
contribution to the estimated  correlation functions, a catalogue with
$100\times$ more random points than galaxies was also created. In the scales
of interest  for our analysis, i.e.   2-25\,h$^{-1}$Mpc, the estimated
errors  were similar  in both  cases.  For  computation  efficiency we
choose to  use the random  catalogue with 20  times as many  points as
real data.

\subsection{AGN sample}

The X-ray selected AGN sample is compiled from the XMM-XXL survey (PI:
Pierre), which covers a total  of about $\rm 50\,deg^2$ split into two
nearly  equal area  fields. The  equatorial subregion  of  the XMM-XXL
overalps with the Canada-France-Hawaii Legacy Survey (CFHTLS) W1 field
and extends to about $\rm 25\,deg^2$ the area covered by the original $\rm
11\,deg^2$  XMM-LSS survey  \citep{Clerc2014}.  The X-ray  survey
layout is presented in Fig. \ref{fig_layout}.

The reduction of  the XMM observations, the construction  of the source
catalogue  and the identification  of the  X-ray sources  with optical
counterparts  follows the  steps described  in 
  \cite{Georgakakis_Nandra2011}.
Specific  details on  the  analysis  of the  XMM-XXL  survey data  are
presented  by  Liu et  al.   (2015, in prep.).   In  brief, the  X-ray  data
reduction is carried  out using the XMM Science  Analysis System (SAS)
version  12.  XMM-XXL  data  observed  prior to  23  January 2012  are
analysed. At  that date the XMM-XXL programme  was partially complete.
As  a result  the  final  catalogue of  the  equatorial XMM-XXL  field
presented in this paper is  missing about $\rm 5\deg^2$ worth of X-ray
data.  This incomplete data coverage is manifested by the white stripes
in Fig.  \ref{fig_layout}.  The  {\sc epchain} and {\sc emchain} tasks
of  {\sc  sas}  are employed  to  produce  event  files for  the  EPIC
\citep[European Photon  Imaging Camera;][]{Struder2001, Turner2001} PN
and  MOS  detectors  respectively.   Periods  of  elevated  background
because  of flares  are identified  and excluded  using  a methodology
similar to that described  by \cite{Nandra2007}.  Sources are detected
independently in five energy bands ($0.5-8$, $0.5-2$, $2-8$, $5-8$ and
$7.5-12$\,keV)  by  applying  a  Poisson false  detection  probability
threshold of $P<4\times10^{-6}$.  Systematic errors in the astrometric
positions  of the  X-ray  sources  are corrected  for  using the  {\sc
eposcorr}  task  of {\sc  sas}  and  adopting  as reference  the  SDSS
\citep[Sloan   Digital    Sky   Survey;][]{Gunn2006}   DR8   catalogue
\citep{Aihara2011}.   Source  fluxes   are  estimated  by  assuming  a
power-law  X-ray  spectrum with  $\Gamma=1.4$,  i.e.   similar to  the
diffuse  X-ray  Background,   absorbed  by  the  appropriate  Galactic
hydrogen    column   density    derived   from    the   HI    map   of
\cite{Kalberla2005}. The  energy to  flux conversion factors  are such
that the counts from the  0.5-2, 0.5-8, 2-8, 5-8 and 7.5-12\,keV bands
are  transformed  to fluxes  in  the  0.5-2,  0.5-10, 2-10,  5-10  and
7.5-12\,keV bands respectively.  The  final catalogue consists of 8,445
unique sources  detected in at least  one of the  above spectral bands
over a total  area of about $\rm 20\,deg^2$.   These X-ray sources are
matched to the SDSS-DR8 photometric catalogue \citep{Aihara2011} using
the Maximum-Likelihood method \citep{Sutherland_and_Saunders1992}.  We
assign  secure  counterparts to  sources  with  Likelihood Ratio  $\rm
LR>1.5$.  At  that cut the  spurious identification rate is  about 6\%
and the total number of optical counterparts is 4,076.

\begin{figure}
\begin{center}
\includegraphics[height=1.\columnwidth]{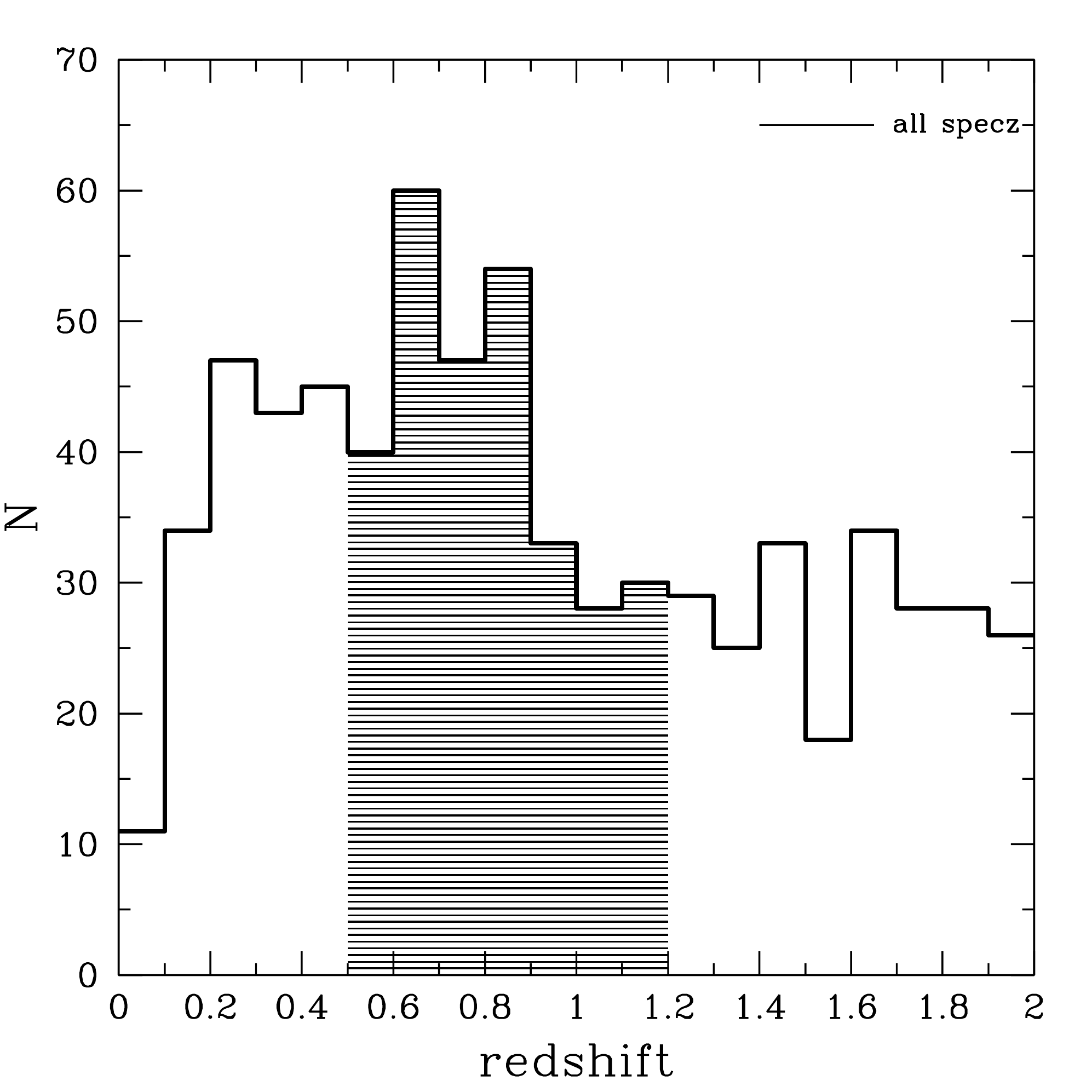}
\end{center}
\caption{The solid histogram shows the redshift distribution of the 813 X-ray sources with
  secure redshifts that lie within the VIPERS spatial mask. The shaded
  region corresponds to the the 318 X-ray sources in the redshift interval
  $0.5<z<1.2$ that are used in the clustering analysis presented in
  this paper.}
\label{fig_xxl_zhist}
\end{figure}

\begin{figure}
\begin{center}
\includegraphics[height=1.\columnwidth]{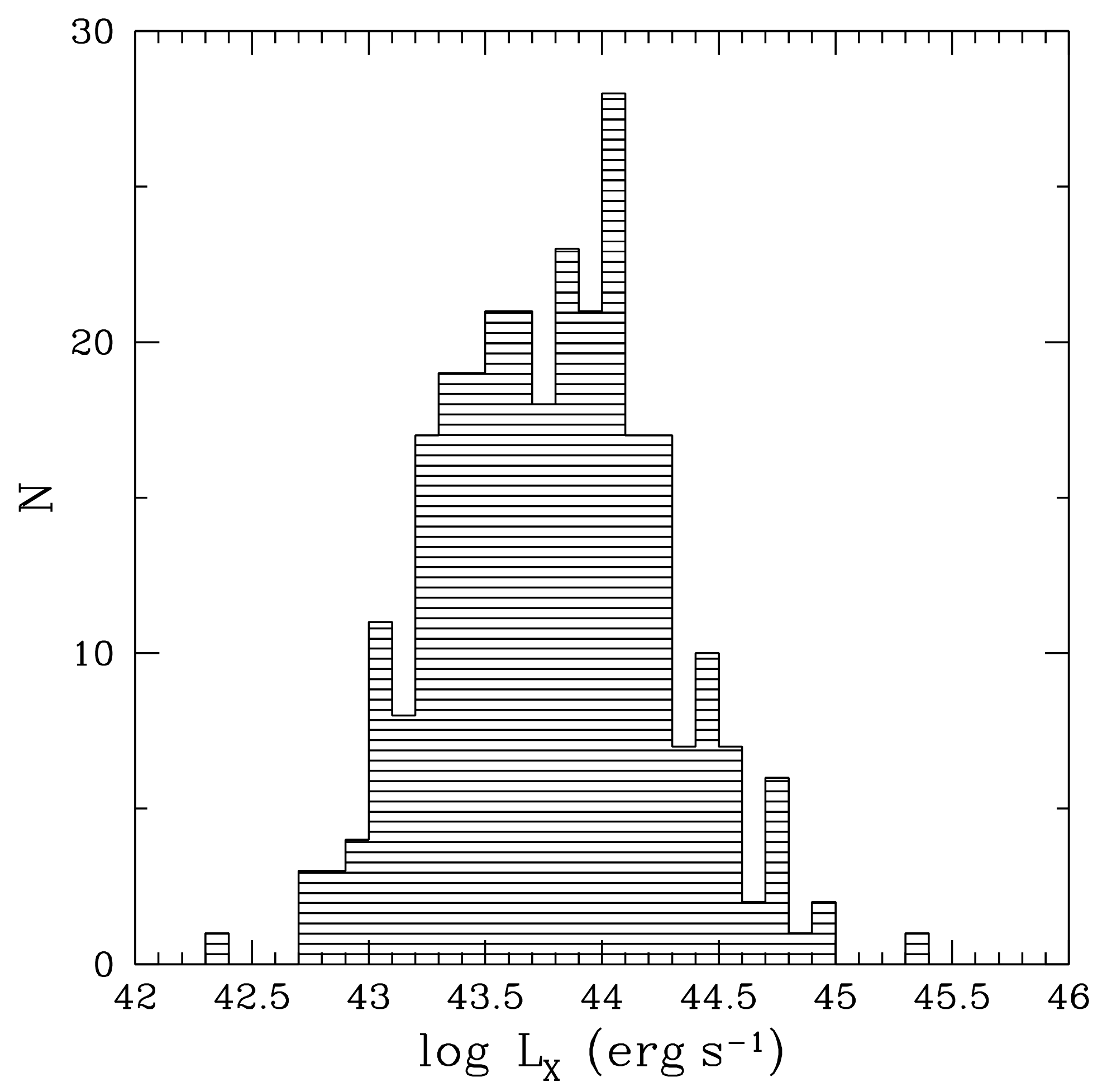}
\end{center}
\caption{The X-ray luminosity distribution of the X-ray selected AGN
  sample in the redshift interval $0.5<z<1.2$ that is used in the
  clustering analysis presented in this paper. The median 
  luminosity of the sample is $\log L_X (\rm  2-10\,keV)=43.6\, erg \,s^{-1}$.}
\label{fig_xxl_lxhist}
\end{figure}

Spectroscopic     redshifts    are     mainly     from    the     SDSS
\citep{Eisenstein2011,    Smee2013}.    In   addition    to   redshift
measurements  obtained as  part of  the SDSS-III's  Baryon Oscillation
Spectroscopic   Survey   \citep[BOSS;][]{Dawson2013}  programme,   the
equatorial  region of  the XMM-XXL  field  was also  targeted by  five
special  SDSS  plates dedicated  to  follow-up  spectroscopy of  X-ray
sources  as   part  of  the  Ancillary  Programs   of  SDSS-III.   The
distribution of  these five plates within the  equatoral XMM-XXL field
is  shown in Fig.   \ref{fig_layout}.  Targets  were selected  to have
$f_X(\rm  0.5-10\,keV) >  10^{-14} \,  erg \,  s^{-1} \,  cm^{-2}$ and
$15<r<22.5$,  where $r$ is  either the  PSF magnitude  in the  case of
optical unresolved  sources (SDSS type=6)  or the model  magnitude for
resolved   sources.   Specific    details   on   these   spectroscopic
observations, including  spectral classification and  redshift quality
flags based in  visual classification are presented by  Menzel et al.
(2015,   in  prep).These  spectroscopic   data  are   complemented  by
additional redshift measurements  from the \cite{Stalin2010} catalogue
and   the  VIPERS  Data   Release  1   \citep{Guzzo2014,  Garilli2014}
spectroscopic  catalogue.  In  total 2816  X-ray sources  are assigned
redshifts and  813 of these are  within the VIPERS  mask. The redshift
distribution   of   the  latter   sample   is   presented  in   Figure
\ref{fig_xxl_zhist}.  In  this paper we limit  the clustering analysis
to  X-ray sources  in the  redshift interval  $0.5<z<1.2$. This  is to
allow  cross-correlation   with  the  VIPERS   galaxy  sample.   X-ray
luminosities are estimated in  the 2-10\,keV band assuming a power-law
X-ray  spectrum with  $\Gamma=1.4$ for  the k-corrections.   The X-ray
luminosity distribution of the $0.5<z<1.2$ sample is presented in Fig.
\ref{fig_xxl_lxhist}.  The median is  $\log \, L_X (\rm 2-10\,keV)=
43.6^{+0.4}_{-0.4}\,erg\,s^{-1}$. The errors correspond to the 16th and 84th
percentiles around the median.

\section{Methodology  and results}\label{section:method} In this paper we infer  the large scale bias of X-ray selected AGN
using   the  2-point  correlation   function.   The   Halo  Occupation
Distribution                     model                     \citep[HOD,
e.g.][]{Peacock2000,Berlind_Weinberg2002}  is often used  to interpret
the 2-point  correlation functions of extragalactic  sources and infer
their large  scale bias as well  as their distribution  in dark matter
halos  \citep[e.g.][]{Padmanabhan2009, delaTorre2013, Richardson2013}.
In this framework, the correlation function at small scales (typically
$\la1-2$\,Mpc)  is  parametrised  by  the  1-halo  term  of  the  HOD,
i.e. pairs of  objects that live in the same  halo, whereas the 2-halo
term, i.e.   pairs of objects  that reside in different  halos, models
the  correlation function  at large  scales ($\ga  2$\,Mpc).  Accurate
measurements of the correlation function at small scales are essential
to  properly  model the  1-halo  term of  the  HOD  and determine  the
distribution  of  AGN  in  dark  matter halos.   This  measurement  is
challenging however, particularly in  the case of the auto-correlation
function of X-ray selected AGN. This is because of the sparsity of AGN
samples  and the typically  small size  of present-day  X-ray surveys,
which  translate to  small  count statistics.  One  can overcome  this
limitation by taking advantage of the higher space density of galaxies
to sample the cosmic web and then determine the distribution of AGN on
it by measuring the  AGN/galaxy cross-correlation function. Shot noise
is suppressed when counting AGN/galaxy  pairs and the impact of sample
variance on the results can be better controlled.  Nevertheless, using
the HOD  to model  the AGN/galaxy cross-correlation  function requires
large  samples   of  both  AGN  and  galaxies   with  well  controlled
systematics. This approach is  therefore still limited to few datasets
\citep[e.g.][]{Miyaji2011,Krumpe2012,Richardson2013},  while there are
also concerns that the HOD  parameterisation of AGN samples may suffer
degeneracies that lead to  ambiguities in the inferred distribution in
dark matter halos \citep[e.g.][]{Shen2013}.

In this paper we  use only the linear regime of  the dark matter power
spectrum to infer  the large scale bias of AGN  and determine the mean
dark matter halo of  the population \citep[e.g][]{Allevato2011, Allevato2012,
  Allevato2014}.    The following sections describe   how    the    AGN/galaxy
cross-correlation function and galaxy auto correlation functions are measured 
and how they are modelled to infer the clustering amplitude of X-ray selected AGN.

\subsection{Spatial clustering measurements via 2-point
  correlation functions}

The equations presented next, are valid for the estimation of both the autocorrelation and cross-correlation function. If an equation needs to be modified to be used for a cross-correlation measurement, then the modified version of the equation is also given. The distance $r$  between two objects in real-space  can be decomposed
into separations along  the line of sight, $\pi$,  and perpendicular to the line
of  sight, $\sigma$.   If $s_1$  and $s_2$  are the  distances  of two
objects  1, 2, measured  in redshift-space,  and $\theta$  the angular
separation between them, then $\sigma$ and $\pi$ are defined as

\begin{equation} 
\pi=(s_2-s_1), $ along the line-of-sight$,
\end{equation}

\begin{equation}
\sigma=\frac{(s_2+s_1)}{2}\theta , $ across the line-of-sight$.
\end{equation}

\noindent The 2-parameter  redshift-space correlation function is then
estimated as

\begin{equation}
\xi(\sigma,\pi) =\frac{DD(\sigma,\pi)}{DR(\sigma,\pi)}-1,
\label{eqn:w_sp}
\end{equation} 

\noindent where $DD(\sigma,\pi)$ are the data-data pairs at separations $\sigma,\pi$. $DR(\sigma,\pi)$ are the AGN-random pairs (cross-correlation) or galaxy-random pairs (galaxy autocorrelation). This estimator has the advantage that requires a random catalogue that accounts only for the selection function of galaxies, which is typically a spatial filter. More advanced estimators, such as  that of \cite[]{LZ1993}, require the construction of random catalogues for the X-ray sources as well. This might introduce systematic biases into the calculations since X-ray observations have variable sensitivity across the field of view, which is challenging to quantify accurately.
In redshift-space the clustering is affected on small scales
by the rms  velocity dispersion of AGN along the line  of sight and by
dynamical  infall of  matter into  higher density  regions.   To first
order, only  the radial component of $\xi(\sigma,\pi)$  is affected by
redshift-space  distortions.  We  can  therefore remove  this bias  by
integrating along the line of sight, $\pi$, to calculate the projected
cross-correlated function, $w_p(\sigma)$:

\begin{equation}
w_p(\sigma)=2\int_0^\infty \xi(\sigma,\pi)d\pi.
\label{eqn:wp}
\end{equation}

\noindent The value of the upper limit of the integral in equation
$\ref{eqn:wp}$, {$\pi_{max}$, has to be determined for the
  estimation of the galaxy autocorrelation and the AGN/galaxy
  cross-correlation functions. For that, the projected correlation
  function, $w_p(\sigma)$, is computed for different $\pi_{max}$
  values and the bias is then estimated for each measurement. Fig. \ref{fig:pimax}
  presents the bias results as a function of $\pi_{max}$. The clustering signal saturates for $\pi_{max}=30$\,Mpc and
  $\pi_{max}=20$\,Mpc, for the galaxy autocorrelation function (stars)
  and the AGN/galaxy cross-correlation function (circles),
  respectively. In each case, the signal is underestimated at smaller scales  and at larger scales the noise from uncorrelated pairs is increased. Therefore, we adopt these two different  $\pi_{max}$ values for each correlation function in our analysis.


\noindent The  real-space  correlation  function  is the Fourier Transform of the linear power spectrum,  $P_{2h}(k)$, i.e.


\begin{equation}
\xi_{DM}^{2h}(r)=\frac{1}{2\pi^2}\int P^{2h}(k) \frac{sin(kr)}{kr}k^2 dk,
\label{eqn:xi_dm}
\end{equation}

\noindent where k is the wavelength of the Fourier Transform and r the distance in real-space. The 2-halo term of the power spectrum can be approximated by \citep{Cooray2002}

\begin{equation}
P^{2h}(k)\approx b^2P_{lin}(k).
\label{eqn:lin_power}
\end{equation}
\noindent Equivalently,

\begin{equation}
\xi^{2h}(r)=b^2\xi_{DM}^{2h}(r),
\label{eqn:proj_dm}
\end{equation}
\noindent or, in terms of the projected correlation function

\begin{equation}
w_{p}^{2h}(\sigma)=b^2w_{DM}^{2h}(\sigma).
\label{eqn:proj_dm}
\end{equation}


\noindent $w_{p}^{2h}(\sigma)$ is the projected correlation function of the extragalactic population under consideration and $w_{DM}^{2h}(\sigma)$ is the projected correlation function of dark matter. Equation \ref{eqn:xi_dm} is estimated, following \cite{Hamana2002} and then the projected correlation function of dark matter is calculated using

\begin{equation}
w_{DM}^{2h}(\sigma)=2 \int _\sigma^\infty \frac{r\xi_{DM}^{2h}(r)dr}{\sqrt{(r^2-\sigma^2)}}
\end{equation}

\noindent The best-fit bias in equation $\ref{eqn:proj_dm}$ is estimated by
applying a $\chi^2$ minimization,
$\chi^2=\Delta^{T}M_{cov}^{-1}\Delta$, where $M_{cov}^{-1}$ is the inverse of the covariance matrix, that quantifies the degree of correlation between the different bins of $w_p(\sigma)$. $\Delta$ is
defined as $w_{p,2h}-w_{p,model}$, where
$w_{p,model}=b^2w_{DM}^{2h}(\sigma)$. In the case of a galaxy autocorrelation measurement, in equation $\ref{eqn:lin_power}$, $b=b_g$, i.e. the galaxy bias, whereas in the case of an AGN/galaxy cross-correlation, $b=b_{AG}$, i.e. the AGN/galaxy bias. The AGN bias can then be inferred via

\begin{equation}
b_{AGN}=\frac{b_{AG}^2}{b_g}
\end{equation}





\noindent For  the  estimation  of the  DMHM we  adopt  the
ellipsoidal   model    of   \cite{Sheth2001}   and    the   analytical
approximations of \cite{Bosch2002}, which  assume that on large scales
the halo mass is only dependent on bias.

The  uncertainties of  the projected
auto-correlation  and  cross-correlation  measurements  are  estimated
using the Jackknife methodology \citep[e.g.][]{Ross2008}.  The area of
the  survey is  split into  $N_{JK}=20$ sections  and  the correlation
function  is measured  $N_{JK}$ times  by excluding  one  section each
time.   For  comparison,  bootstrap   errors  are  also  estimated  by
performing  100 resamplings with  replacement \citep[e.g.][]{Loh2008}.
Both  the Jackknife  and  bootstrap resampling  methods yield  similar
covariance matrices and hence,  similar uncertainties for the inferred
bias and the dark matter halo mass.  The errors estimates presented in
the rest of  this paper are based on the  Jackknife method.

\subsection{Applying corrections for redshift incompleteness to the VIPERS galaxy sample}

For the determination of the AGN/galaxy cross-correlation function and
the  galaxy auto-correlation  function in  the XMM-XXL  field equation
(\ref{eqn:w_sp}) has  to be modified to account  for the spectroscopic
window function and the sampling rate of the VIPERS galaxy sample (see
Section 2.1). Following  the analysis of de la Torre  et al. (2013) to
account  for the  completeness  variations from  quadrant to  quadrant
(large-scale  incompleteness), each  galaxy  pair is  weighted by  the
inverse of the effective sampling rate in each quadrant Q

\begin{equation} 
w(Q)=(SSR(Q)\times TSR(Q))^{-1}.
\end{equation}

\noindent  Incompleteness at  small
scales  because of  slit collisions  or the  finite number  of science
slits available  (see Section  2.2) is corrected  for by  adopting the
methodology of \cite{Hawkins2003}.  Each galaxy  pair  separated by  angle $\theta$  is
weighted up by the factor

\begin{equation}
\frac{1}{w^A(\theta)}=\frac{1+w_s(\theta)}{1+w_p(\theta)}.
\label{hawkins}
\end{equation}

\noindent  $w_p(\theta)$ is  the angular  correlation function  of the
parent  sample (potential  targets) and  $w_s(\theta)$ is  the angular
correlation function  of those sources  that have a  measured spectrum
(spectroscopic   sample). The large-scale incompleteness is also included in these calculations, i.e. each  galaxy pair is weighted by  w(Q). The small-scale
incompleteness  affects  the   measurements  on  scales  smaller  than
$\theta=0.03$\,deg  or 1\,h$^{-1}$Mpc  comoving at  $z=0.7$ (Fig.
\ref{fig:corrections}).   This is
smaller   than   the   scales    of   interest   for   our   analysis,
$2-25$\,Mpc. Nevertheless,  the effect has been taken  into account in
the clustering  measurements. The total weight applied on each DD and DR pair is

\begin{equation}
D_GD_G(r)=\sum_{i=1}^{N_G}\sum_{j=i+1}^{N_G}w_i(Q_i)w_j(Q_j)w^A_{GG}(\theta_{ij})
\end{equation}

\begin{equation}
D_gR(r)=\sum_{i=1}^{N_g}\sum_{j=1}^{N_r}w_i(Q_i)
\end{equation}

\noindent Same corrections are applied in the case of the AGN/galaxy cross-correlation measurements.  Fig. \ref{fig:corrections} (triangles) shows the small-scales incompleteness effect, for this case. The total weight assigned in each AGN/galaxy pair is

\begin{equation}
D_AD_g(r)=\sum_{i=1}^{N_A}\sum_{j=1}^{N_g}=w_j(Q_j)w^A_{AG}(\theta_{ij})
\end{equation}

\begin{table*}
\caption{X-ray AGN and galaxy samples in the VIPERS field within the redshift range of $0.5\leq z\leq 1.2$.}
\centering
\setlength{\tabcolsep}{1.5mm}
\begin{tabular}{lcccccc}
       \hline
 \hline
{sample} & {No. of sources} & {$\rm <z>$} &{$\rm <\log L_X>$ }  & {b$_{CCF}$}  & {b$_{ACF}$} &{logM$_{DMH}$}\\
 & & & ($\rm erg \,s^{-1}$) & & & (h$^{-1}$\,M$_\odot$)  \\
       \hline
    \\
AGN  & 318 & 0.81 & 43.6  & {$1.32^{+0.08}_{-0.08}$}   &{$1.43^{+0.18}_{-0.18}$} & {$12.50^{+0.22}_{-0.30}$} \\
\\
galaxies &20,109  & 0.71 & & & {$1.22^{+0.03}_{-0.03}$}  & {$12.28^{+0.05}_{-0.07}$}  \\
		\\
       \hline
\label{table:agn_samples}
\end{tabular}
\end{table*}

\subsection{Results}

Figure \ref{fig:wp_main} plots the projected auto-correlation function
of the  galaxy sample  and the AGN/galaxy  cross-correlation function.
These  are estimated  from equations  (3) and  (4) after  applying the
spatially  variable weights  described in  Section 3.2.   We  then use
equation (8) in combination  with the dark matter correlation function
of \cite{Hamana2002}  to model the measured  correlation functions
and  estimate the bias  of AGN.   In this  calculation we  use spatial
scales in the range $2-25$\,Mpc to fit the correlation functions.  The
inferred bias  is then converted to  mean dark matter  halo mass using
the model by \cite{Sheth2001} as explained in Section 3.1.

For the galaxy sample  we estimate $b=1.22^{+0.03}_{-0.03}$ and a mean
dark  matter  halo mass  of  $\log  M  / (M_{\odot}\,h^{-1})  =  12.28
^{+0.05}_{-0.07}$.   Within   the  uncertainties,  this   is  in  fair
agreement    with   $\log    M   /    (M_{\odot}\,h^{-1})    =   12.36
^{+0.04}_{-0.05.}$ determined by \cite{delaTorre2013} using both the W1
and W4 VIPERS fields.

The estimated bias  and mean dark matter halo of  AGN are presented in
Table \ref{table:agn_samples}. We find  that X-ray selected AGN in the
XMM-XXL  field are associated  with halos  of average  mass $\log  M /
(M_{\odot}\,h^{-1})  =  12.50  ^{+0.22}_{-0.30}$.   For  the  sake  of
reproducibility of these results  the covariance matrices for both the
galaxy auto-correlation and the AGN/galaxy cross-correlation functions
are presented in Table \ref{table:covariance}.

It has been shown \citep{bosch2013} that the integration of equation  \ref{eqn:wp} to a finite $\pi_{max}$, rather than infinity, introduces systematic errors in the calculation of the projected correlation function and may lead to the underestimation of the inferred bias of extragalactic sources because of residual redshift distortions. We investigate the impact of this effect to the results by estimating the correction factors proposed by \cite{bosch2013} and applying them on the galaxy auto-correlation function and the galaxy-AGN cross-correlation function. We use the linear power spectrum in real-space and redshift-space to estimate the correction factor, as described in equation (48) of \cite{bosch2013} and then apply this correction function on the projected correlation function, following their equation (47). This approach yields an AGN bias of $1.53^{+0.20}_{-0.19}$, that corresponds to a halo with average mass $\log  M / (M_{\odot}\,h^{-1})  =  12.63  ^{+0.23}_{-0.28}$. In the rest of the paper we use the values of bias and DMH mass listed in Table  \ref{table:agn_samples}, i.e. not corrected for residual velocity distortions. This is because we are comparing our results to previous studies on the clustering of AGN that have ignored this effect.

\begin{figure}
\begin{center}
\includegraphics[scale=0.43]{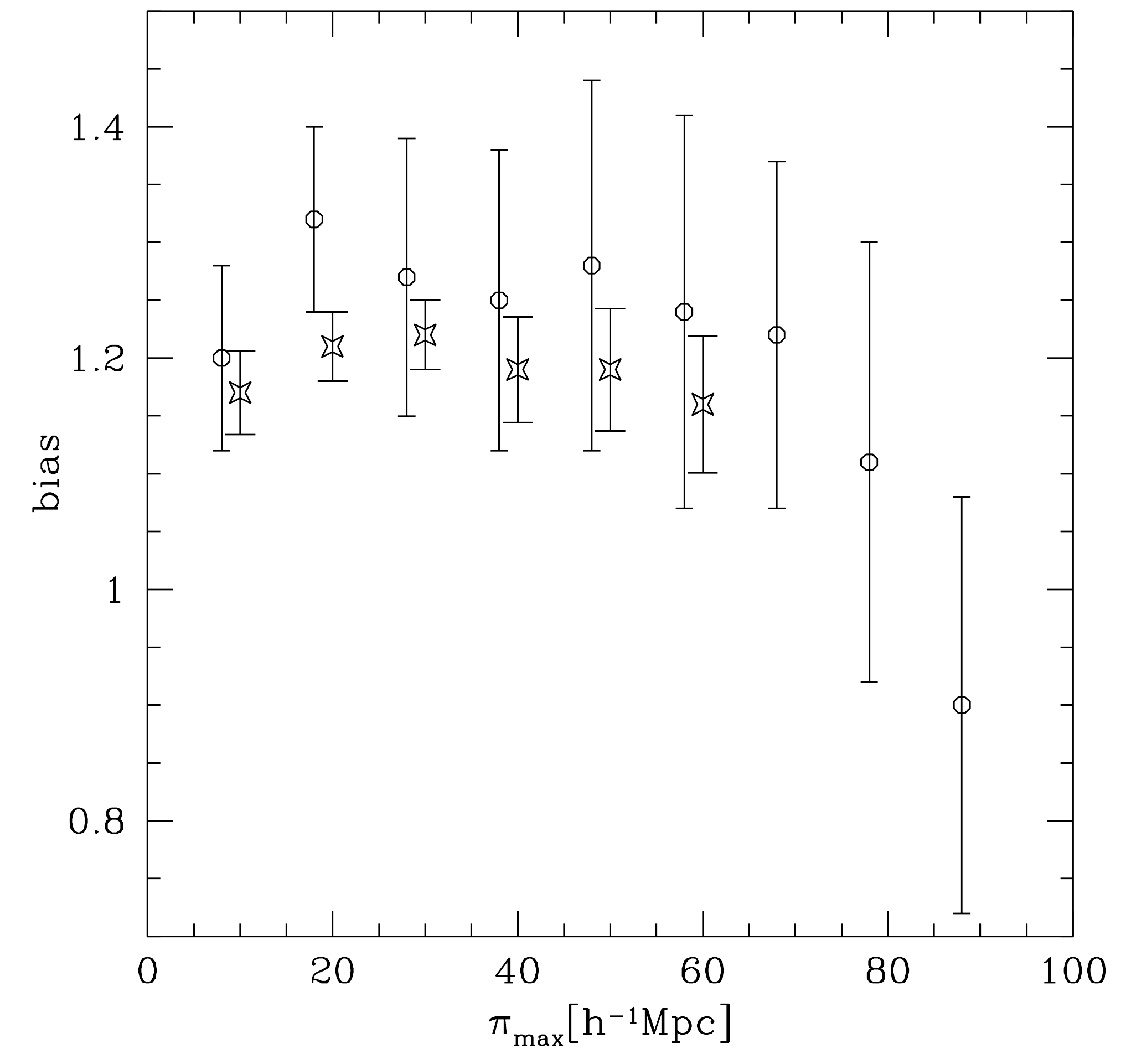}
\end{center}
\caption{Bias estimations for the galaxy autocorrelation (stars) and
  the AGN/galaxy cross-correlation (circles) as a function of
  $\pi_{max}$. The clustering signal saturates for $\pi_{max}=30$\,Mpc
  and $\pi_{max}=20$\,Mpc, for the galaxy autocorrelation function and
  the AGN/galaxy cross-correlation function, respectively. The circles
  are offset in the horizontal direction by -2\,Mpc for clarity.}
\label{fig:pimax}
\end{figure}

\begin{figure}
\begin{center}
\includegraphics[scale=0.43]{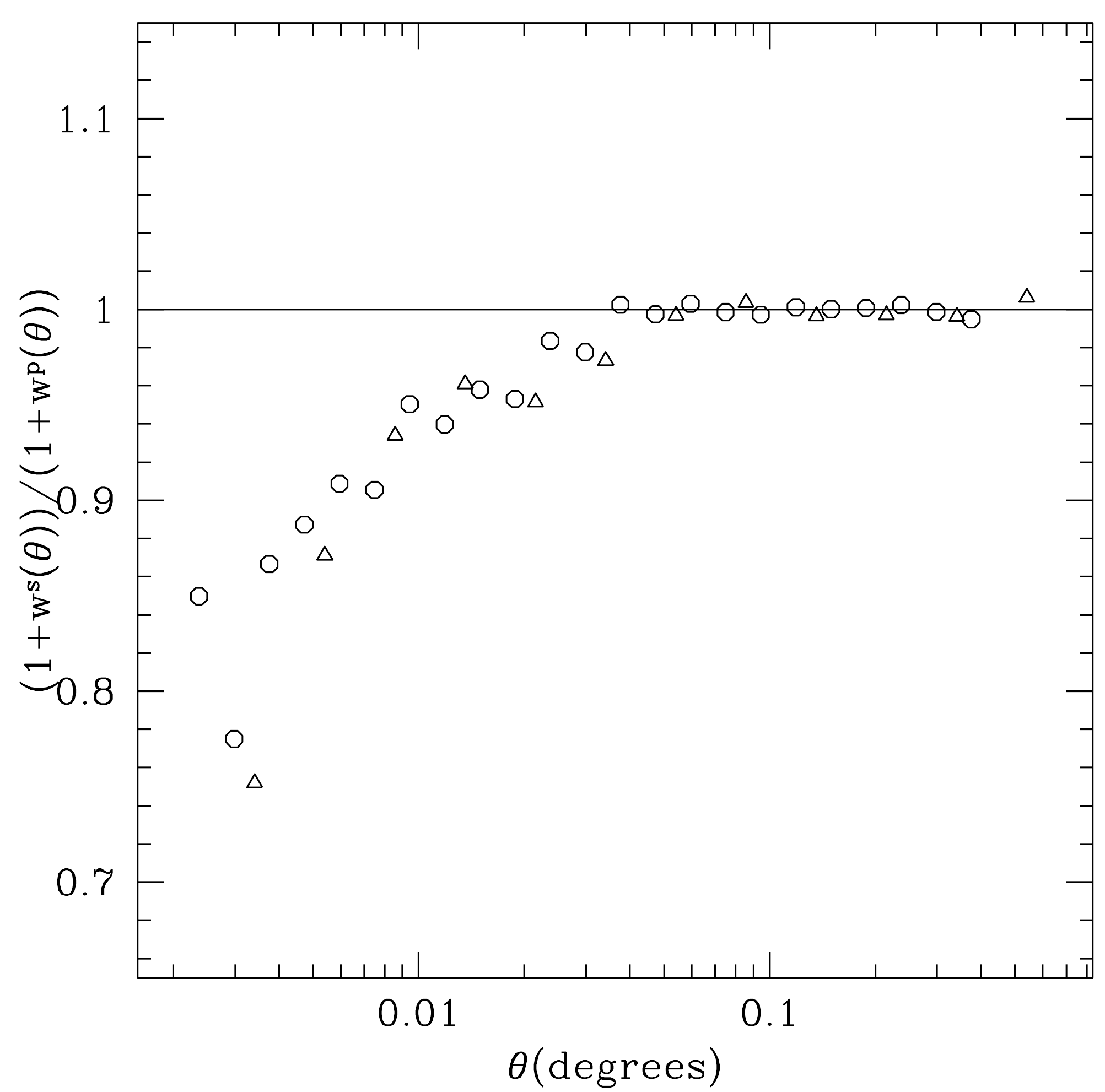}
\end{center}
\caption{The effect of small-scale incompleteness (see text). Circles present the angular pair completeness of the galaxy autocorrelation function and triangles the AGN/galaxy cross-correlation function, measured following Hawkins et al. (2003). The small-scale incompleteness affect the measurements on scales $\theta<0.03$\,deg. The large-scale incompleteness has also been taken into account.}
\label{fig:corrections}
\end{figure}

\begin{figure}
\begin{center}
\includegraphics[scale=0.43]{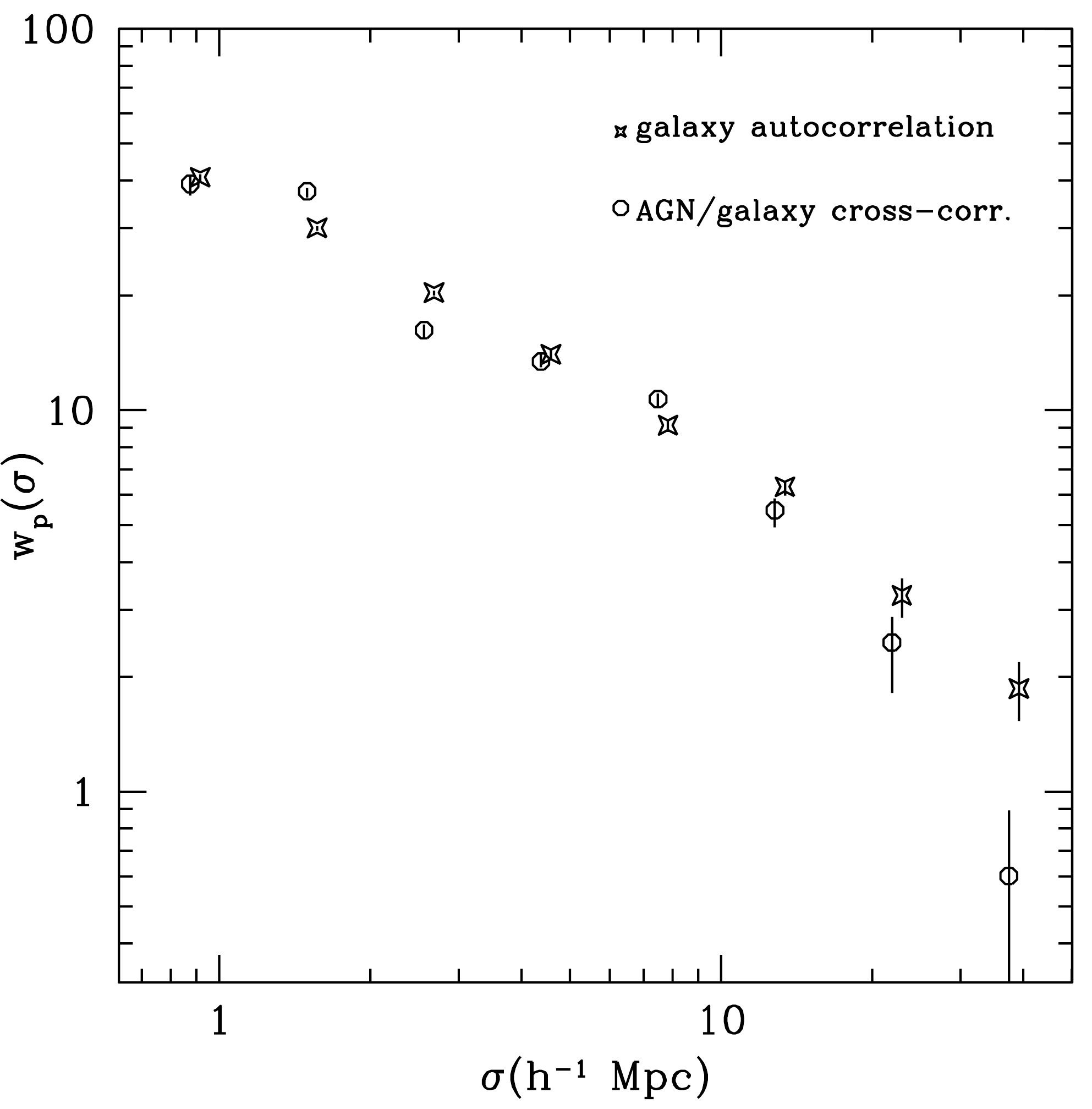}
\end{center}
\caption{The projected galaxy autocorrelation function (stars) and the AGN/galaxy cross-correlation function (circles). The errors are the 16th and 84th percentiles of the distribution of $w_p(\sigma)$ from 20 Jackknife regions. The circles are offset in the horizontal direction by log = -0.02 for clarity.}
\label{fig:wp_main}
\end{figure}

\begin{table*}
\caption{The covariance matrices (not normalized) for the galaxy autocorrelation function and the AGN/galaxy cross-correlation function.}
\centering
\begin{tabular}{lcccccccccccc}
       \hline
\multicolumn{6}{c}{galaxy autocorrelation} && \multicolumn{6}{c}{AGN/galaxy cross-correlation}\\ 
 \hline
{$\sigma$(h$^{-1}$\,Mpc)} & 2.68 & 4.58 & 7.84 & 13.41 & 23.94 &&& 2.68 & 4.58 & 7.84 & 13.41 & 23.94 \\
       \hline
    \\
2.68 &  2.801     &  1.783     &  1.944   &    1.581     & 0.784   &  & &8.667 &       1.925 &       0.072  &  2.053 &   0.8723 \\
\\
4.58  &      &    1.924 &     1.741 &     1.329  &    0.693  && & &     7.615 &      2.099 &       2.240 &       3.245 \\
		\\
7.84  &      & &   1.997 &   1.809 &      1.219  &  &&   &   &       2.956 &  -0.341 &    0.742\\
		\\
13.41  &     &       &     &  2.139 &  1.855  &  & & &      & &  3.922&      2.496\\
		\\
23.94  &     &    &    &       &      2.217 & &   &    &    &    &    &     4.939  \\
		\\
\end{tabular}
\label{table:covariance}
\end{table*}

\section{Discussion}


We use  the overlap  between the XMM-XXL  X-ray survey and  the VIPERS
optical spectroscopic  sample in the  CFHTLS-W1 field to  estimate the
cross-correlation function  of relatively luminous
[$L_X(\rm 2 -  10 \, keV ) \approx 10^{43.6} \,  erg \, s^{-1}$] X-ray
selected AGN  with galaxies in the redshift  interval $z=0.5-1.2$.  We
infer a mean bias for  the AGN $b=1.43^{+0.18}_{-0.18}$ and an average
DMH mass of $\log M / (M_{\odot}\,h^{-1}) = 12.50 ^{+0.22}_{-0.30}$.

A number of  studies determine mean DMH masses  for X-ray selected AGN
at  luminosities  similar  to  those  probed  by  the  XMM-XXL  sample
presented in this paper.  \cite{Krumpe2012}  for example, explore  the clustering of
AGN  in the  RASS \citep[ROSAT  All-Sky  Survey;][]{Voges1999}.  Their
highest  redshift  sub-sample  has  $z=0.36-0.50$  and  a  mean  X-ray
luminosity $\log  L_X (\rm 2-10\,keV) \approx 43.8$  (units erg/s; see
\cite{Fanidakis2013a}  for  the  conversion  of RASS  band  fluxes  to
2-10\,keV luminosities).  For these  sources they estimate a mean dark
matter    halo    mass    of    $\log    M    /    (M_{\odot}\,h^{-1})
=12.51^{+0.28}_{-0.25}$.  \cite{Starikova2011}  measure the clustering
of X-ray  selected AGN  in the Chandra  Bo$\ddot{\textrm{o}}$tes field
\citep{Murray2005}.  Their sample with  $z=1.00-1.68$ has a mean X-ray
luminosity $\log L_X (\rm 2-10\,keV) \approx 43.6$ (units erg/s).  The
inferred  DMH  mass is  $\log  M /(M_{\odot}\,h^{-1})  =12.69\pm0.24$.
Although the  DMH masses  determined in the  above works are  in broad
agreement with our estimates,  the different redshift intervals render
the comparison  hard to interpret.   Additionally, \cite{Allevato2011}
estimate  dark  matter halo  masses  for  X-ray  selected AGN  in  the
XMM-COSMOS  field  \citep{Hasinger2007}   that  are  higher  than  the
measurements above.  For their  subsample with median redshift $z=1.3$
and median X-ray luminosity $\log L_X(\rm 2 - 10 \, keV ) \approx 44.0
\,  erg  \,  s^{-1}$  they  determine $\log  M  /  (M_{\odot}\,h^{-1})
=13.12\pm{0.12}$.   These  higher  dark  matter halo  mass  measurements
compared  to  our  results  and  other  previous  studies  at  similar
redshifts and luminosities may  be related to sample variance effects.
The  COSMOS   field  is  known   to  be  rich  in   cosmic  structures
\citep{Gilli2009, Skibba2015}, which may bias clustering measurements.
Our approach  for inferring the  clustering properties of AGN  via the
AGN/galaxy cross-correlation  function should at  least partly account
for sample variance effects.   Nevertheless, larger samples are needed
to further explore this issue.

The  X-ray  AGN  sample  presented  in  this paper  does  not  span  a
sufficiently large  luminosity baseline  to explore variations  of the
clustering amplitude with luminosity.  We can nevertheless compare our
results with previous studies on the clustering properties of moderate
luminosity [$\log L_X(\rm  2 - 10 \, keV ) \la  43.4\, erg \, s^{-1}$]
X-ray selected  AGN at  redshifts similar to  the sample  presented in
this    paper   \citep[e.g.][]{Coil2009,    Gilli2009,   Starikova2011,
Allevato2012,  Mountrichas2013}.  We restrict  the comparison  to those
clustering studies that  use redshift slices for the  AGN samples that
are relatively narrow and and  similar to ours ($0.5<z<1.2$).  This is
because of  suggestions that the clustering  strength of AGN  may be a
strong   function   of   both   redshift  and   accretion   luminosity
\citep{Fanidakis2013a}.  The investigation of the luminosity dependence
of the AGN clustering therefore  requires samples that are selected at
similar and relatively narrow redshift intervals.

Figure  \ref{fig:galform} plots DMH  mass as  a function  of accretion
luminosity and compares  the mean DMH mass determined  for the XMM-XXL
AGN with  previous estimates in the literature  at lower luminosities.
Moderate luminosity  AGN in this  figure are associated with  mean DMH
masses $\approx  10^{13} \rm \, h^{-1}  \,M_{\odot}$.  These estimates
are about  0.5\,dex more massive than  the mean DMH  mass inferred for
the XMM-XXL  AGN sample.  This  is evidence for a  negative luminosity
dependence of  the AGN  clustering at $z\approx0.8$,  i.e.  decreasing
mean  dark matter  halo mass  with increasing  accretion luminosity.
This  trend  is  contrary  to  previous  claims  for  a  weak  positive
correlation between accretion luminosity  and dark matter halo mass at
$z\approx1$  \citep[e.g.][]{Plionis2008, Krumpe2012, Koutoulidis2013}.
We  attribute this  apparent discrepancy  to the  different luminosity
intervals  of previous  studies.   Figure \ref{fig:galform},  combines
clustering measurements  from both deep small-area  surveys and bright
wide-area  samples  (XMM-XXL)  and   therefore  covers  a  much  wider
luminosity baseline compared to any previous study at $z\approx0.8$.

The  XMM-XXL data  point  in Fig.   \ref{fig:galform}  also links  the
relatively high clustering properties of moderate luminosity X-ray AGN
(i.e.  DMH  masses $\approx 10^{13}  \rm \, h^{-1}  \,M_{\odot}$) with
those of powerful optical/UV  selected QSOs, which typically reside in
DMH   of   few   times  $10^{12}$\,h$^{-1}M_\odot$   \citep{Croom2005,
  daAngela2008,   Ross2009}.   Variations   of  the   clustering  with
accretion luminosity may  indicate different AGN triggering mechanisms
\citep[e.g.][]{Allevato2011,  Mountrichas2013} and/or  different black
hole fuelling  modes \citep[e.g.][]{Fanidakis2013b, Fanidakis2013a} as
a  function of  AGN luminosity.   The latter  scenario is  explored in
Figure \ref{fig:galform}, which overplots  the predictions of the {\sc
  galform}   semi-analytic   model  \citep{Bower2006,   Fanidakis2012,
  Fanidakis2013a} on the DMH  mass vs $L_X(\rm 2-10\,keV)$ plane.  The
{\sc galform} model postulates two modes for growing black holes.  The
first is  associated with star-formation  events (``starburst'' mode).
It  assumes that  a  fraction of  the  cold gas  in  galaxies that  is
available  for  star-formation accretes  onto  the  SMBH and it occurs when the host galaxy experiences a disk instability, or a major/minor merger in a gas rich disk.  The  second
fuelling mode  of {\sc galform}  is decoupled from  star-formation and
takes place  in quiescent galaxies  (``hot-halo'' mode).  This  is the
case of  DMHs with  a diffuse hot  gas component  in quasi-hydrostatic
equilibrium, which  accretes onto the SMBH without  being cooled first
onto the galactic  disk.  By construction the two  black hole fuelling
modes in {\sc galform} operate in different large scale envrironments.
Starburst  AGN are  typically found  in relatively  small  DMHs, while
hot-halo mode AGN  also extend to very massive  haloes \citep[for more
  details see][]{Fanidakis2013a}.  The interplay between the two modes
produces the  complex relation  between accretion luminosity  and mean
DMH  mass shown  in  Figure \ref{fig:galform}.   At  low and  moderate
luminosities  both   modes  co-exist.   There  is   therefore  a  wide
distribution  of DMH  masses, including  some massive  ones associated
with  hot-halo  mode AGN,  which  bias the  average  DMH  mass of  the
population  to  high  values. At  high  accretion
luminosities however,  the ``hot-halo'' mode  becomes sub-dominant and
the mean DMH  mass of the population drops sharply  to the typical DMH
mass of the Starburst mode AGN.  The observational constraints plotted
in  Figure \ref{fig:galform}, including  the XMM-XXL  data-point, also
suggest an  abrupt change in the  mean dark matter halo  mass of X-ray
selected  AGN,  in  qualitative   agreement  with  the  {\sc  galform}
predictions. At  the  same  time  however,  the  comparison  of  the
observations  with  the {\sc  galform}  SAM  predictions also  reveals
important differences.  The transition  luminosity, where the mean DMH
mass of the  AGN population shows a sudden drop,  is brighter by about
0.5\,dex  in the model  compared to  the data.   In {\sc  galform} the
accretion  luminosity  at which  the  starburst  mode fueling  becomes
dominant  over the  hot-halo mode  accretion (and  therefore  the mean
clustering  properties of  the population  change) is  related  to the
adopted value for the viscosity ($\alpha=0.087$ in {\sc galform}), the
cooling properties of the parent DMH and the timescale that the cooled
gas is accreted onto the  cenrtal SMBH in the hot-halo accretion mode.
Modifying  these parameters  could potentially  improve  the agreement
between the observed  clustering properties of X-ray AGN  and the {\sc
  galform} predictions.  Such modifications however,  may also require
retuning of other parameters of {\sc galform}.

Alternatively,  the  apparent  discrepancy between  the  observational
results  and the  {\sc  galform} model  predictions  on the  turnover
luminosity  in  Figure \ref{fig:galform}  may  indicate that  physical
processes other  than hot  gas accretion are  responsible for  low and
moderate    luminosity   X-ray    selected    AGN   at    $z\approx1$.
\cite{Ciotti2007}  for example  suggest  that mass  loss from  evolved
stars could  provide a sufficient  supply of gas to  trigger recursive
accretion events onto supermassive black holes in early-type galaxies.
Population studies  at low redshift  indeed suggest that  this process
may be responsible for the fuelling of low and moderate luminosity AGN
in  passive  hosts  \citep{Kauffmann_Heckman2009}.   These  early-type
galaxies are  also expected to  be associated on average  with massive
dark  matter  halos  \citep[e.g.][]{Zehavi2011,  Mostek2012},  thereby
increasing the  mean bias  of the AGN  population at low  and moderate
accretion luminosities.  Another  possibility is that of gravitational
interactions  between galaxies  or  tidal disruption  events in  dense
environments.   These processes  may also  play an  important  role in
fuelling AGN activity at low and moderate luminosities.

All the scenarios outlined above require a baseline black hole fuelling
mode that operates in small dark  matter halos and produces AGN over a
wide luminosity  baseline (including  powerful ones) and  an additonal
process that takes  place in denser environemnts and  produces low and
moderate  luminosity   accretion  events.   In  this   picture  it  is
variations in  the width  of dark matter  halo mass distribution  at a
given accretion luminosity that produces the observed luminosity trend
of the mean AGN dark matter halo mass in Figure \ref{fig:galform}.

The considerations above also highlight the importance of constraining
the full  distribution of  AGN in  DMHs, not just  the mean,  which is
sensitive    to    the     tails    of    a    skewed    distribution.
\cite{Mountrichas2013} for example, find  that the average dark matter
halo mass of moderate luminosity X-ray selected AGN drops from $\log M
/ ( M_{\odot}  \, h^{-1} ) \approx  13.0$ to $\log M /  ( M_{\odot} \,
h^{-1}  ) \approx  12.5$ after  removing only  5 per  cent of  the AGN
population associated  with X-ray selected groups.  This suggests that
most moderate luminosity X-ray AGN live in halos with masses few times
$10^{12}$\,h$^{-1}M_\odot$.  \cite{Leauthaud2015}   assumed  that  AGN
hosts follow  the same  stellar--to--halo mass relation  as non-active
galaxies \cite[see  also][]{Georgakakis2014, Krumpe2015} and propose  that the DMH
distribution  of  moderate-luminosity  X-ray  AGN at  $z<1$  peaks  at
relatively  low  masses with  a  tail  extending  to group-size  halos
($\approx10^{13}$\,h$^{-1}M_\odot$). Constraints    on   the   Halo
Occupation Distribution  of AGN are broadly  consistent these findings
\citep[e.g.][]{Miyaji2011,  Allevato2012}. The  analysis  presented in
this paper, suggests that at high accretion luminosities [$L_X(\rm 2
- 10 \, keV ) \approx  10^{43.6}\, erg \, s^{-1}$], the high mass tail
of the X-ray AGN DMH distribution  is reduced and as a result the mean
DMH mass is considerably lower than moderate luminosity AGN.

\begin{figure}
\begin{center}
\includegraphics[scale=0.42]{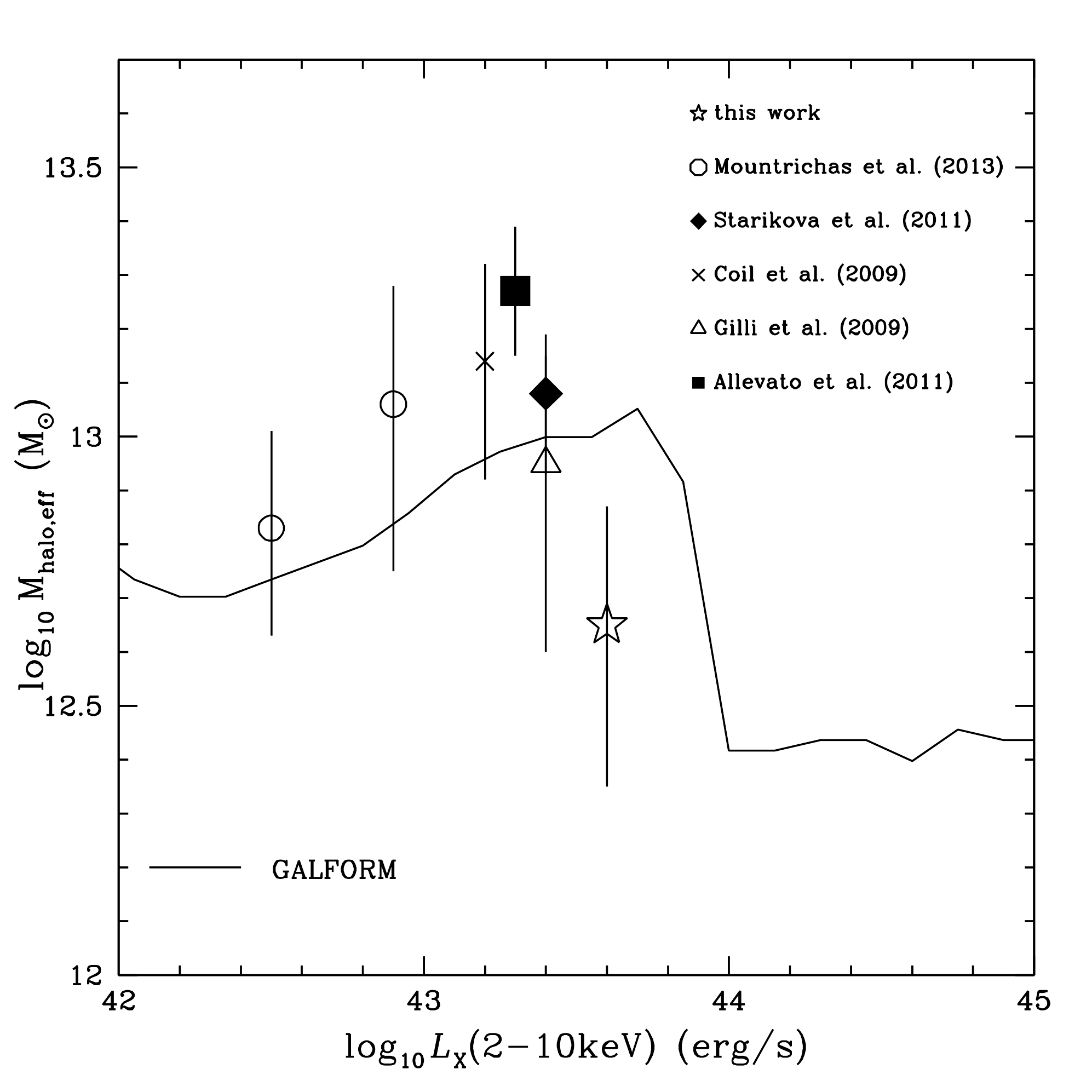}
\end{center}
\caption{DMH mass as a function of accretion luminosity. Moderate luminosity AGN are associated with mean DMH masses of $\approx10^{13}$\,h$^{-1}$\,M$_{\odot}$, i.e. about 0.5 dex more massive than the mean DMH mass inferred for the XMM-XXL AGN sample. Halo mass calculations are taken from Table 1 in Fanidakis et al. (2013) and have been converted to the same cosmology as the GALFORM model, i.e. H$_0$=70\,Km\,s$^{-1}$\,Mpc$^{-1}$. The solid line presents the complex relation between luminosity and halo mass, predicted by the {\sc galform}  model. The predictions of the model at the $L_X>10^{44}$\,erg/s, corresponds to the clustering level of the starburst mode. This mode has only a weak dependence on the X-ray luminosity and therefore can be extrapolated to fainter luminosities.}
\label{fig:galform}
\end{figure}

\section{Conclusions}

A  clustering  analysis is  applied  to relatively luminous [$\log  L_X  (\rm
2-10\,keV)\approx  43.6$\,erg/s] X-ray  AGN in  the  redshift interval
$z=0.5-1.2$  extracted from  the  $\sim \rm25\,deg^2$  of the  XMM-XXL
field. Only scales of the correlation function that include the linear
regime of the power spectrum are  modelled, to derive the AGN bias and
DMHM of  the host  galaxies. The analysis  reveals that  these sources
live in  haloes of $\log M/(M_{\odot}\,h^{-1})=12.50^{+0.22}_{-0.30}$.
This  mass is  lower than  moderate  luminosity X-ray  AGN at  similar
redshifts  ($\log   M/(M_{\odot}\,h^{-1})\approx13$)  and  similar  to
UV/optical selected QSOs.  This is evidence for a decreasing mean dark
matter halo mass for  AGN with increasing accretion luminosity.  These
results are consistent with suggestions that the dark matter halo mass
distribution of  AGN is broad  and includes a massive-end  tail, which
skews measurements  of the mean dark  matter halo mass of  AGN to high
values.   This tails  appears to  become sub-dominant  with increasing
accretion luminosity.  We  also discuss the results in  the context of
cosmological semi-analytic models in  which the broad dark matter halo
mass distributon of AGN is  related to different fuelling modes of the
central  balck hole.   We show  that the  observed  negative accretion
luminosity dependence  of the X-ray  AGN clustering is  in qualitative
agreement with such models.

\section{Acknowledgments}

The authors are grateful to the anonymous referee for helpful comments and Takamitsu Miyaji for useful discussions. GM  acknowledgments financial  support from  the {\sc  thales} project
383549  that is jointly  funded by  the European  Union and  the Greek
Government in the framework  of the programme ``Education and lifelong
learning''.  This paper uses data from the VIMOS Public Extragalactic
Redshift Survey (VIPERS). VIPERS has been performed using the ESO Very
Large Telescope, under the "Large Programme" 182.A-0886. The
participating institutions and funding agencies are listed at
http://vipers.inaf.it. Funding for  SDSS-III  has been  provided  by the  Alfred
P.  Sloan  Foundation, the  Participating  Institutions, the  National
Science  Foundation,  and the  U.S.  Department  of  Energy Office  of
Science. The  SDSS-III web site is  http://www.sdss3.org/. SDSS-III is
managed by the Astrophysical Research Consortium for the Participating
Institutions of the SDSS-III Collaboration including the University of
Arizona,  the  Brazilian   Participation  Group,  Brookhaven  National
Laboratory,  Carnegie Mellon  University, University  of  Florida, the
French  Participation Group, the  German Participation  Group, Harvard
University,  the Instituto  de Astrofisica  de Canarias,  the Michigan
State/Notre Dame/JINA  Participation Group, Johns  Hopkins University,
Lawrence  Berkeley  National  Laboratory,  Max  Planck  Institute  for
Astrophysics, Max  Planck Institute for  Extraterrestrial Physics, New
Mexico State  University, New York University,  Ohio State University,
Pennsylvania  State University,  University  of Portsmouth,  Princeton
University,  the  Spanish Participation  Group,  University of  Tokyo,
University  of Utah,  Vanderbilt University,  University  of Virginia,
University of Washington, and Yale University.

\bibliography{mybib}{}

\bibliographystyle{mn2e}

\end{document}